\documentclass[journal]{IEEEtran}
\ifCLASSINFOpdf
\else
\fi
\usepackage{amsfonts}
\usepackage{amsmath}
\usepackage{graphicx}
\usepackage{color}
\usepackage{multirow}
\usepackage{soul}
\usepackage{pdflscape}
\usepackage{lscape}
\usepackage{rotating}
\usepackage{subfig}
\usepackage{booktabs}
\usepackage{placeins}
\usepackage{xr}
\usepackage{amsthm}
\usepackage{bbding}
\usepackage{array}
\usepackage{relsize}
\newcommand{\Var}{\operatorname{Var}} 
\newcommand{\E}{\operatorname{E}} 
\usepackage{cite}

\usepackage{esint}
\usepackage[colorlinks=true, allcolors=blue]{hyperref}
\usepackage{arydshln}
\usepackage{eurosym}
\usepackage{float}
\usepackage{algorithm,algorithmic}
\newcommand{\SWITCH}[1]{\STATE \textbf{switch} (#1)}
\newcommand{\ENDSWITCH}{\STATE \textbf{end switch}}
\newcommand{\CASE}[1]{\STATE \textbf{case} #1\textbf{:} \begin{ALC@g}}
\newcommand{\ENDCASE}{\end{ALC@g}}

\newcommand{\DEFAULT}{\STATE \textbf{default:} \begin{ALC@g}}
\newcommand{\ENDDEFAULT}{\end{ALC@g}}
\newcommand{\DEFAULTLINE}[1]{\STATE \textbf{default:} }

\usepackage[font={scriptsize,it}]{caption}

\begin{document}

%
\title{Real-Time Sensor Anomaly Detection and Recovery in Connected Automated Vehicle Sensors}

\author{Yiyang~Wang,
        Neda~Masoud, Anahita Khojandi,~\IEEEmembership{Member,~IEEE}

\thanks{\copyright 2020 IEEE. Personal use of this material is permitted. Permission from IEEE must be obtained for all other uses, in any current or future media, including reprinting/republishing this material for advertising or promotional purposes, creating new collective works, for resale or redistribution to servers or lists, or reuse of any copyrighted component of this work in other works. doi: 10.1109/TITS.2020.2970295.\newline
Yiyang Wang and Neda Masoud are with the University of Michigan, Ann Arbor, MI 48109,  USA (e-mail: yiyangw@umich.edu,  nmasoud@umich.edu). Anahita Khojandi is with the University of Tennessee at Knoxville, TN 37996, USA  (email:khojandi@utk.edu).}}
\markboth{Accepted to be Published in: IEEE Transactions on Intelligent Transportation Systems}%
{Shell \MakeLowercase{\textit{et al.}}: Bare Demo of IEEEtran.cls for IEEE Journals}
%

\maketitle

\begin{abstract}
In this paper we propose a novel observer-based method to improve the safety and security of connected and automated vehicle (CAV) transportation. The proposed method combines model-based signal filtering and anomaly detection methods. Specifically, we use adaptive extended Kalman filter (AEKF) to smooth sensor readings of a CAV based on a nonlinear car-following motion model. Under the assumption of a car-following model, the subject vehicle utilizes its leading vehicle's information to detect sensor anomalies by employing previously-trained One Class Support Vector Machine (OCSVM) models. This approach allows the AEKF to estimate the state of a vehicle not only based on the vehicle's location and speed, but also by taking into account the state of the surrounding traffic. A communication time delay factor is considered in the car-following model to make it more suitable for real-world applications. Our experiments show that compared with the AEKF with a traditional $\chi^2$-detector, our proposed method achieves a better anomaly detection performance. We also demonstrate that a larger time delay factor has a negative impact on the overall detection performance. 

\end{abstract}


%
\IEEEpeerreviewmaketitle

\section{Introduction}\label{sec:sec_1}
 Numerous studies within the past decade have focused on connected and automated vehicle (CAV) technology, which is considered to be an integral part of the future of the intelligent transportation system (ITS) field \cite{shladover2018connected}. CAVs have the potential to transform the ITS field by introducing numerous safety, mobility, and environmental sustainability benefits \cite{abdolmaleki2019itinerary,van2019path,abdolmaleki2019vehicle,masoud2017autonomous}.
 A CAV system combines connected vehicle (CV) and automated vehicle (AV) technologies, creating a synergistic impact that goes well beyond the benefits that each of these technologies can offer in isolation. 
 It is envisioned that CAVs, with diverse degrees of connectivity and automation, will lead the path toward the next generation of transportation systems, which is more intelligent, efficient, and sustainable \cite{meyer2014road,litman2017autonomous}.

CAVs use wireless technologies to enable communication and cooperation not only among vehicles but also between vehicles and the transportation infrastructure. By using dedicated short-range communication (DSRC) \cite{kenney2011dedicated}, or other types of communication technologies, vehicles and roadside units (RSUs) are able to continuously transmit and receive information such as speed, position, acceleration, braking status, traffic signal status, etc., through what is called a Basic Safety Message (BSM). These communication messages have a range of about 400 meters and can detect high-risk situations that may not be observable otherwise due to traffic, terrain, or weather \cite{kenney2011dedicated}. The CAV technology  extends and enhances currently available crash avoidance systems that use radars and cameras to detect collision threats by enabling CAVs to warn their surrounding vehicles of collisions and potentially hazardous circumstances. In addition, they provide  mobility and sustainability benefits by enabling platoon formation, which can increase road capacity and reduce fuel consumption.
However, as vehicles and infrastructures become more interconnected and automated, the vulnerability of their components to faults and/or deliberate malicious attacks increases. This vulnerability is exacerbated by the increase in vehicle-to-vehicle (V2V) and vehicle-to-infrastructure (V2I) communications, which increase a vehicle's external connection interfaces. At the system level, CAVs and the infrastructure can be viewed as individual nodes in a large interconnected network, where a single anomaly or malicious attack can easily propagate through this network, affecting other network components (e.g., other vehicles, traffic control devices, etc.). Therefore, there is an increasing demand for cyber security solutions, e.g., anomaly detection methods, in CAV sensor systems to enhance safety and reliability of CAVs and the entire network.

Anomaly detection in CAV sensors is an important but also challenging task. 
A traveling CAV could use the most recent history of data to detect anomalies. Presence of an anomaly in the pattern of data collected from a CAV sensor system can imply ($i$) a subset of sensors are faulty, or ($ii$) there has been a malicious attack. In both cases, it is vital to detect the anomalies and exclude the anomalous data from the decision making process. 

An anomaly detection scheme introduces two types of errors -- false negatives and false positives. It is easy to see that a false negative error can allow falsified data to affect trajectory planning, which could lead to fatal consequences. Although less apparent, a false positive error can have consequences that are just as severe. Consider a situation where an actual event in the network (e.g., an unexpected braking from a downstream vehicle) has led to an abrupt change in the pattern of observed data. If the vehicle falsely detects such an unexpected change as a fault/attack and discards the information, it may lead to the CAV not reacting to such abrupt changes in the network appropriately and in a timely manner, creating dangerous, and potentially fatal, scenarios. In order to prevent this type of false positive error, it is necessary for vehicles to incorporate network-level information in their anomaly detection scheme.

In addition to distinguishing between real changes in network conditions and anomalies, the anomaly detection methods should be able to identify the noise introduced by sensors and the communication channel, exacerbated by potential communication delay, as well as the missing values in the collected data. Moreover, due to resource constraints for each vehicle, the anomaly detection techniques in CAVs need to be lightweight, and implementable in real-time.

Anomalous sensor behavior could manifest itself in various forms and representations. Several faulty sensor behaviors are discussed in \cite{sharma2010sensor}. Petit et al. \cite{petit2015potential} summarize the taxonomy of intrusions or attacks on automated vehicles, among which the false injection attack is considered to be the most dangerous attack.  In this paper, we consider five types of the anomalous sensor behavior resulting from both sensor faults and false injection attacks. We base this paper on the sensor failure and/or attack taxonomy provided by Sharma et al. \cite{sharma2010sensor} and Van et al. \cite{van2019real}:

\begin{enumerate}
 
\item Short: A single, sharp and abrupt change in the observed data between two successive sensor readings.

\item Noise: An increase in the variance of the sensor readings. Different from Short, the Noise anomaly type occurs across \textit{multiple} successive sensor readings. 

\item Bias: A temporarily constant offset from the sensor readings.

\item Gradual drift: A small and gradual drift in observed data during a time period. Over time, a gradual drift can result in a large discrepancy between the observed and the true state of the system.  

\item Miss: Lack of available data during a time period.
\end{enumerate}

In this paper, we do not explicitly account for `miss,' which can result from DoS attacks preventing the exchange of information. However, note that  `miss,' depending on its duration, {\color{black}can be viewed as `short' or `bias'}, where the sensor reading is non-existent instead of showing a wrong value. Hence, it can partially be addressed using the same methods for detecting `short' or `bias'. For examples of specific scenarios that could lead to these anomalies, refer to \cite{ni2009sensor}.

In order to successfully detect different types of anomalies, avoid falsely identifying unexpected changes in the network as anomalies, and mitigate the impact of random noise/missing values, we develop a novel and comprehensive framework that combines the adaptive extended Kalman Filter (AEKF) with a car following motion model, and employs a data-driven fault detector. Our framework is capable of accounting for delay in observing the environment, introduced by a congested communication channel and/or delayed sensor observation. Specifically, we use a car-following model to govern the motion of the vehicle in order to capture the interaction between the subject vehicle and its immediate leading vehicle. We demonstrate that the time delay incorporated in the motion model renders the traditional $\chi^2$ fault detector \cite{brumback1987chi} not appropriate for anomaly detection, and propose and implement a One Class Support Vector Machine (OCSVM) model for anomaly detection \cite{scholkopf2001estimating} together with AEKF instead. We demonstrate the power of the proposed framework in detecting various types of anomalies. 

Our main objective in this study is to detect sensor anomalies and to recover the corrupt signals by utilizing the surrounding vehicles' information. To this end, the following assumptions are made: 

\begin{enumerate}
    
    \item Vehicles move according to a car-following model (i.e., under adaptive cruise control mode), have access to location and velocity of their leader (either through BSMs or using their on-board sensors), and are able to control over their own acceleration rates.
    
    \item A known time delay (e.g., communication, sensing, and/or reaction delay) is applied to the input vector of the car-following model.

\end{enumerate}

The rest of the paper is organized as follows: Section II provides a brief review of the existing related work in the field of anomaly detection in CAVs. Section III introduces the formulation of the problem and our method. 
In section IV we conduct a case study based on a well-known car-following model. Finally, in section V, we conclude this paper.

\section{Related Work}

Anomaly detection research has generated a substantial volume of literature over the past few years, as it is an important and challenging problem in many disciplines including but not limited to automotive systems \cite{muter2011entropy,muter2010structured}, wireless networks \cite{rajasegarar2008anomaly}, and environmental engineering \cite{hill2007real,hill2010anomaly}. Anomaly detection methods are used in a variety of applications including fault diagnosis, intrusion detection, and monitoring applications. In some cases if the source of an anomaly can be quickly identified, appropriate reconfiguration control actions can be made in order to avoid or minimize potential loss.

In the past few years, a variety of methods have been developed to detect anomalous behavior, and/or identify the source of anomaly \cite{isermann1984process,hwang2010survey}. Examples of anomaly detection methods include observer-based methods \cite{clark1975detecting,wunnenberg1987sensor}, parity relation methods \cite{deckert1977f,gertler1997fault}, and parameter estimation methods \cite{baskiotis1979parameter}, etc. Among them, observer-based (quantitative model-based) fault detection is a common fault detection approach, as discussed in \cite{hwang2010survey}. Observer-based fault detection is based on the residual (or innovation) sequence obtained from using a mathematical model and (adaptive) thresholding. 
In this paper we study anomaly detection in CAVs using observer-based anomaly detection. 

Anomalous sensor behavior in CAVs could result from both sensor failures or malicious cyber attacks. Sensor readings may be influenced by a variety of factors, leading to collection of faulty information \cite{checkoway2011comprehensive,realpe2015sensor,pous2017intelligent}. For example, environmental perturbations and sensor age may result in higher probabilities of failure. A short circuit, loose wire connection, or low battery supply are among other reasons that may cause inaccurate data reporting, including an unexpectedly high variation of sensor reading, or noise \cite{ni2009sensor}. 

Additionally, malicious attacks may cause anomaly in sensor readings. CAVs have several internal and external cyber attack surfaces through which they can be accessed and compromised by ill-intended actors  \cite{petit2015potential,checkoway2011comprehensive,koscher2010experimental,weimerskirch2015overview,yan2016can}. Petit and Shladover  \cite{petit2015potential} showed that false injection of information and map database poisoning are two of the most dangerous potential attacks on CAVs.  For example, the infrastructure (i.e., RSU) or a neighboring vehicles can transmit fake messages (e.g., WAVE Service Advertisement, BSM), which may in turn generate wrong, and potentially harmful, reactions (e.g., spurious braking), placing CAV occupants and other road users in life-threatening situations. There are several existing studies that illustrate the vulnerability of CAV sensors, e.g., speed, acceleration and location sensors, to cyber attacks or faults. For in-vehicle speed and acceleration sensors, a false injection attack mentioned in \cite{petit2015potential} through the CAN bus or the on-board diagnostics (OBD) system could induce any of the four types of anomalies considered in this paper. As another example, Trippel et al. demonstrates that an acoustic injection attack could lead to anomalous sensor values for the in-vehicle acceleration sensor \cite{trippel2017walnut}. Lastly, for the location measurement from the GPS, both the operating environment of the vehicle and GPS spoofing/jamming attacks may result in anomalous sensor values \cite{faughnan2013risk}. Note that in this study we only consider false injection attacks whose manifestations can be described by four types of anomaly defined in Section I. As such, the study leaves out any types of attacks that do not impact sensor readings.

\looseness-1 Despite the severe consequences of failing to detect sensor anomalies in CAVs, there is a scarcity of anomaly detection techniques in the ITS literature. Only a limited number of studies have focused on cyber security in CAVs, or more generally in ITS. In \cite{park2015sensor}, Park et al. use graph theory based on a transient fault model to detect transient faults in CAVs. Christiansen et al. \cite{christiansen2016deepanomaly} combine background subtraction and convolutional neural networks to detect anomalies/obstacles. Muter et al. \cite{muter2011entropy} and Marchetti et al. \cite{marchetti2016evaluation} use entropy-based methods to detect anomalies (attacks) in in-vehicle networks. Faughnan et al. measure the discrepancy between redundant sensor readings to detect hijacking in unmanned aerial vehicles \cite{faughnan2013risk}. 
van Wyk et al. \cite{van2019real} use a CNN - Kalman Filter - $\chi^2$-detector hybrid method to detect and identify sensor anomaly in a CAV system. 

In this paper, we focus on detection of anomalous sensor readings and recovery of the corrupt signals. We propose an observer-based anomaly detection method, which combines a well-known filtering technique, namely, AEKF, to smooth the CAV sensor values, and a machine learning method, i.e., OCSVM, to learn the normal vehicle behavior, with the objective of detecting anomalous behavior. Specifically, we utilize a car following model to take into account the information from the leading vehicle, so as to better detect anomalies by reducing the false positive error rate. Additionally, to make our methodology robust to practical network conditions and improve its anomaly detection performance, we account for time delay in perceiving the environment, which could arise from communication delay or sensor observation delay. 

One of the major differences of this paper with our past work is that in \cite{van2019real}  we examine multiple sensor readings for each type of sensor at the same time by feeding multiple sensor readings into a CNN network. However, in this paper, for each type of sensor we rely on readings from a single sensor only and  propose a novel anomaly filtering and detection technique accordingly. Another major difference is that this work takes into account the state of the leading vehicle when conducting anomaly detection for the subject vehicle. These two major differences make the two frameworks fundamentally different, and applicable to different scenarios. Finally, in this paper we have replaced the traditional $\chi^2$-detector,  which was used in \cite{van2019real}, with a OCSVM model. Our experiments show that by cooperating leading vehicle's information and using OCSVM, we achieve a better detection performance compared to the traditional $\chi^2$-detector. 
To the best of our knowledge, this is the first study that detects CAV sensor anomaly by utilizing leading vehicle's information, i.e., by incorporating a car-following model into a continuous state-space model with time delay. Additionally, given the fact (demonstrated in the paper) that the model noise does not follow a Gaussian distribution, rendering the traditional $\chi^2$ test inapplicable, we propose an OCSVM model in order to deal with the bias and abnormal distribution of innovation caused by time delay.

\section{METHODS}

In this section, we first discuss how a car-following model with time delay can be used to describe the motion (also known as state-transition) model in AEKF. Next, we formulate a new continuous nonlinear state-space model with discrete measurement based on a car-following motion model. The continuous state-transition model represents the intrinsic nature of a vehicle's response to the actions of its immediate downstream traffic, and the discrete measurement model represents the mechanics of sensor sampling, as is the case in practice. Based on the proposed state-space model, we propose an anomaly detection method,  which combines AEKF and OCSVM. Also a traditional $\chi^2$-detector is discussed and its performance is compared with that of OCSVM.

\subsection{Car-Following Model with Time Delay}
Consider the car-following model in \cite{treiber2014traffic}:

\begin{equation}\label{eq:1}
\begin{aligned}
d_n(t) &= x_{n-1}(t)-x_n(t)\\
\dot{d}_n(t) &= v_{n-1}(t) - v_n(t)\\
\dot{v}_{n}(t) &= f(v_{n}(t-\tau),d_n(t-\tau), \dot{d}_n(t-\tau))\\
\end{aligned}
\end{equation}
where $\dot{v}_{n}(t)$, $v_n(t)$, $x_n(t)$ are respectively the acceleration, speed, and location of the $n$th vehicle, to which we refer as the `subject vehicle', and  $d_n(t)$ and $\dot{d}_n(t)$ are the distance gap and the speed difference between the subject vehicle and its leading vehicle, the $(n-1)$th vehicle, respectively. Parameter $\tau$ denotes time delay, also known as the `perception-reaction time', i.e., the period of time lapsed from the moment the leading vehicle performs an action, to the moment the subject vehicle executes an action in response. 
Function $f$ is the stimulus function. 

$\dot{v}_{n}(t)$ in Equation (\ref{eq:1}) can be recast in the following form:

\begin{equation}\label{eq:2}
\dot{v}_{n}(t) = {\color{black}q}\left(x_{n}(t-\tau),v_{n}(t-\tau),x_{n-1}(t-\tau),v_{n-1}(t-\tau))\right.
\end{equation}
where $q$ produces the same output as $f$, given  a  different set of inputs.

We define a state vector in continuous time as:
\begin{equation}
    \label{eq:state}
    s_n(t) = \left[x_{n}(t),v_{n}(t)\right]^T \in \mathbb{R}^{2},
\end{equation}
where $x_{n}(k)\in \mathbb{R}$ and $v_{n}(k)\in \mathbb{R}$. Note that, without loss of generality, $x$ and $v$ can be extended to vector form to allow for incorporating historical location and speed observations, respectively, into the state-space model, when desired. 

Recasting equation (\ref{eq:2}) as a function of $s_n(t)$ produces a car following model that maps the state into an actionable decision for the subject vehicle: 
\begin{equation}\label{eq:2.1}
\begin{aligned}
\dot{v}_{n}(t) 
:= &f_{vc}\left(s_n(t-\tau),u_{n}(t-\tau))\right.
\end{aligned}
\end{equation}
where $u_{n}(t) = [x_{n-1}(t),v_{n-1}(t)]^T$ is the input vector containing information received from the leading vehicle, and  $f_{vc}$ denotes the stimulus function describing velocity in a continuous sate space.

\subsection{Continuous State-Discrete Measurement State-Space Model}
We now define a state-space model with a continuous state-transition model and discrete measurements. Using previous definition of the state vector $s_n(t)$, the state-transition model satisfies the following differential equation:

\begin{equation}\label{eq:11}
\begin{aligned}
\dot{s}_n(t) &= 
\begin{bmatrix} 
\dot{x}_n(t)  \\
\dot{v}_n(t)  
\end{bmatrix} \\
&= 
\begin{bmatrix}
e_2^T s_n(t)\\
f_{vc}(s_n(t-\tau),u_{n}(t-\tau))
\end{bmatrix}.\\
\end{aligned}
\end{equation}
where $e_2 = [0,1]^T$.

When $\tau = 0$, the state-space model in equation (\ref{eq:11}) satisfies the Markovian property, allowing for applying AEKF. However, in practice, a variety of factors including time required for data processing and computations as well as delays in the communication network can cause $\tau$ to be non-zero. As such, in practice AEKF cannot be applied to equation (\ref{eq:11}), since the derivative of the state vector is determined by multiple previous state vectors. 

In order to apply AEKF, we approximate equation (\ref{eq:11}) in the following way: We assume the acceleration of each vehicle is bounded within the interval $[a_{min},a_{max}]$, where $a_{min}\leq 0$ and $a_{max} > 0$ indicate the magnitude of the maximum deceleration and acceleration rates, respectively. Based on the assumption of bounded acceleration, we can obtain lower and upper bounds on the approximation of $v_n(t)$:
$$e_2^T s_n(t-\tau) + a_{min}\tau \leq v_n(t)\leq e_2^T s_n(t-\tau) + a_{max}\tau.$$

Then a delay differential equation (DDE), describing the delayed state-transition model, can be used to approximate equation (\ref{eq:11}):

\begin{equation}\label{eq:12}
\begin{aligned}
\dot{s}_n(t) &= 
\begin{bmatrix} 
\dot{x}_n(t)  \\
\dot{v}_n(t)  
\end{bmatrix} \\
&= 
\begin{bmatrix}
e_2^T s_n(t-\tau) + \int_{t-\tau}^{t}a_n(r)dr\\
f_{vc}(s_n(t-\tau),u_{n}(t-\tau))
\end{bmatrix}\\
&\approx 
\begin{bmatrix}
e_2^T s_n(t-\tau)\\
f_{vc}(s_n(t-\tau),u_{n}(t-\tau))\\
\end{bmatrix}\\
&= g_{sc}(s_n(t-\tau),u_{n}(t-\tau))
\end{aligned}
\end{equation}
where $a_n(t)$ is the acceleration of the $n$th vehicle at time $t$, and $g_{sc}$ denotes the derivative of the continuous-time state space.
Finally, we obtain a continuous-time state-transition model with discrete-time measurement as the following:
\begin{equation}\label{eq:state space}
\begin{aligned}
\dot{s}_n(t) &= g_{sc}(s_n(t-\tau),u_{n}(t-\tau))+\theta(t)\\
z_n(t_k) &= h(s_n(t_k))+\eta(t_k),\ k \in \{0 \cup \mathbb{Z}^+\}\\
\end{aligned}
\end{equation}
where $h(\cdot)$ is the measurement function, $z_n(\cdot)$ denotes sensor reading of the $n$th vehicle, $\theta(t)$ and $\eta(t_k)$ are the process noise and the observation noise, respectively, which are assumed  to be mutually independent, $t_{k+1} = t_k + \Delta t, ~k \in \{0 \cup \mathbb{Z}^+\}$, and $\Delta t$ is the sampling time interval for sensors. Note that $\theta(t)$ accounts for the error introduced by the approximation steps in equation (\ref{eq:12}). 

\subsection{Adaptive Extended Kalman Filter with Fault Detector}

Extended Kalman Filter (EKF) is a well-established method used for timely and accurate estimation of the dynamic state of a non-linear system \cite{wan2006sigma}. One important issue that needs to be addressed in EKF is how to properly set up the covariance matrices of process noise (i.e., $Q$) and measurement noise (i.e., $R$). The performance of EKF is highly affected by proper tuning of $Q$ and $R$ \cite{mohamed1999adaptive}, while in practice these parameters are usually unknown a priori. Therefore, we apply an adaptive extended Kalman filter (AEKF) to approximate these matrices. 

An EKF is used to estimate state vector $s_n(t_k)$ from sensor reading $z_n(t_k)$.  Let $\hat{s}(k|k-1)$ and $P(t_k|t_{k-1})$ denote the state prediction and state covariance prediction at time $t_k$, given the estimate at time $t_{k-1}$, respectively. Note that for ease of notation, we omit subscript $n$. Hence, considering the state-space model in equation (\ref{eq:state space}), the EKF consists of the following 3 steps:

\emph{Step 0 - Initialize:}
To initialize EKF, the mean values and covariance matrix of the states are set up at $k=0$ as the following:
\begin{equation} \label{eq:7}
\begin{aligned}
    &\hat{s}_{k|k-1} = \E[s(t_0)]\\
    & P_{k|k-1} = \Var[s(t_0)].\\
\end{aligned}
\end{equation}

\emph{Step 1 - Predict:}
The state and its covariance matrix at $t_{k-1}$ are projected one step forward in order to obtain the a priori estimates at time $t_k$,

\begin{equation} \label{eq:7.1}
    \begin{aligned}
    \textrm{Solve}
    &\begin{cases}
        \dot{\hat{s}}(t) = g_{sc}(\hat{s}(t-\tau),u(t-\tau)),\\
        \dot{P}(t) = F(t-\tau)P(t-\tau)+P(t-\tau)F(t-\tau)^T + Q(t)\\
    \end{cases}\\
    \textrm{with}
    &\begin{cases}
        \hat{s}(t_{k-1}) = \hat{s}_{k-1|k-1}\\
        P(t_{k-1}) = P_{k-1|k-1}
    \end{cases}\\
    \Rightarrow 
    &\begin{cases}
        \hat{s}_{k|k-1} = \hat{s}(t_k)\\
        P_{k|k-1} = P(t_k)
    \end{cases}
    \end{aligned}
\end{equation}
where $F(t-\tau) = \frac{\partial g_{sc}}{\partial s}|_{\hat{s}(t-\tau),u(t-\tau)}$ is the first-order approximation of the Jacobian matrix of function $g_{sc}(\cdot)$.

\emph{Step 2 - Update:}
\begin{equation} \label{eq:8}
\begin{aligned}
    &\nu_k = z(t_k) - h(\hat{s}_{k|k-1})\\
    &S_k = H(t_k) P_{k|k-1}H(t_k)^T + R_k\\
    &K_k = P_{k|k-1}H(t_k)^T S_k^{-1}\\
    &\hat{s}_{k|k} =\hat{s}_{k|k-1}+ K_k\nu_k\\
    &P_{k|k} = P_{k|k-1} - K_k H(t_k)P_{k|k-1}\\
\end{aligned}
\end{equation}
where 
$H(t_k) = \frac{\partial h}{\partial s}|_{\hat{s}_{k|k-1}}$, $Q(t)$ is the covariance matrix of the process noise at time $t$, $R_k = R(t_k)$ is the covariance matrix of the measurement noise at time $t_k$, and $\nu_k$ is innovation (i.e., the difference between the measurement and the prediction) at time $t_k$. 

Since in practice $Q$ and $R$ are usually unknown, based on the work in \cite{akhlaghi2017adaptive} with slight modifications, we apply an AEKF to estimate these two matrices by using a moving estimation window of size $M$, as follows:
\begin{equation}\label{eq:10}
    \begin{aligned}
        \mu_k &= z(t_k) - h(\hat{s}_{k|k}) \\
        \hat{R}_k &=
        \sum_{i = 1}^{M} \lambda_{i}\left(\mu_{k-i+1}~ \mu_{k-i+1}^T\right. \\
        & \left.+ H(t_{k-i+1})P_{k-i+1|k-i+1}H(t_{k-i+1})^T\right) \\
        \hat{Q}_k &=\sum_{i = 1}^{M} \lambda_i K_{k-i+1} \nu_{k-i+1} \nu^{T}_{k-i+1}  K_{k-i+1} \\
    \end{aligned}
\end{equation}
where $\mu_k$ is the residual at time $t_k$, which is the difference between actual measurement and its estimated value using the information available at time $t_k$, \{$\lambda_i, i = 1,2,...,M\}$ are forgetting factors, and $\sum_{i = 1}^{M}\lambda_i = 1$. Note that using a moving window, as we place more weight on previous estimates, less fluctuation of $\hat{R}_k$ and $\hat{Q}_k$ will incur, and it takes longer for the model to capture changes in the system. Additionally, note that we replace $Q(t)$ in equation (\ref{eq:7.1}) with $\hat{Q}_k$ during the time interval $[t_{k-1},t_k]$.

One of the traditional fault detectors used in conjunction with Kalman filter is the $\chi^2$-detector \cite{brumback1987chi,bar1995multitarget,geng2008adaptive}. Since AEKF is a special type of Kalman filter, the $\chi^2$-detector can be seamlessly applied to AEKF as well. Specifically, it constructs $\chi^2$ test statistics to determine whether the new measurement falls into the gate region with the probability determined by the gate threshold $\sigma$, as shown in the following:
\begin{equation}
\begin{aligned}
V_{\gamma}(k) = \{z:&(z - \hat{z}_{k|k-1})^T S_k^{-1} (z-\hat{z}_{k|k-1})\leq \sigma\}.\\
\end{aligned}
\end{equation}
where $\hat{z}_{k|k-1}$ is the predicted value of measurement at time $t_k$.
The $\chi^2$ test statistics for the fault detector is defined as
\begin{equation}
\chi^2(t_k)= \nu_k^T S_k^{-1}\nu_k.
\end{equation}

\begin{figure}[t] 
\centering
\includegraphics[width=0.45\textwidth]{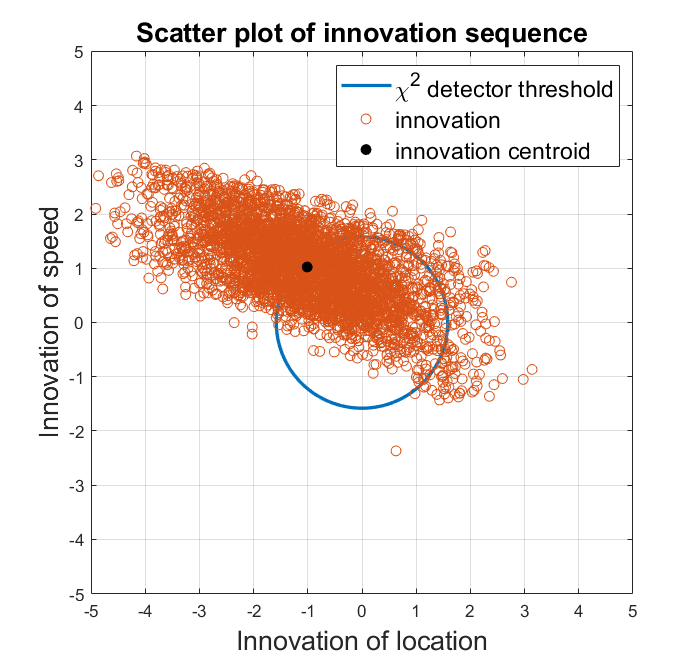}
\caption{An example where the normalized innovation sequence is not zero mean due to time delay and imperfect model, where $\tau = 0.5$ second. The threshold $\sigma$ of $\chi ^2$-detector is 2.5.}
\label{fig:non0innovation}
\end{figure}

For the $\chi^2$ test to provide meaningful results, the innovation $\nu_k$ should be zero mean Gaussian distributed with covariance $S_k$. However, in reality, the innovation can follow a non-zero mean Gaussian distribution if there is bias in the background (e.g., due to non-zero mean process noise or imperfect model), as shown in figure \ref{fig:non0innovation}. This figure displays a scatter plot of normalized innovation generated from the training dataset in our experiments, when there exits a time delay and $\int_{t-\tau}^{t}a_n(r)dr$ in equation (\ref{eq:12}) is not zero. Moreover, in practice the variance of the normalized innovation is not {normally distributed} when the noise does not follow a Gaussian distribution. In the 2-dimensional case, the $\chi^2$-detector defines a circular boundary with its center located at $(0,0)$, i.e., the blue lines in figure \ref{fig:non0innovation}, which corresponds to the thresholding boundary of the $\chi^2$-detector with $\sigma =2.5$. Therefore, in order to correctly detect anomalies, the $\chi^2$-detector requires the data to be zero mean and normally distributed. In such a scenario, the $\chi^2$-detector would not be a good detector since it will generate higher rates of false positives and false negatives. As such, the boundary should be shifted toward the true mean in order to achieve a `fair' boundary on both sides.

In the context of our problem, the approximation introduced in equation (\ref{eq:12}) can generate such a bias. This approximation assumes the value $\int_{t-\tau}^{t}a_n(r)dr$ to be zero. As such, unless the acceleration and deceleration rates during the period $[t~,~ t+\tau]$ sum to zero, or the time delay $\tau$ is equal to zero, the resulting bias would degrade the performance of the $\chi^2$-detector.
Consequently, we propose a novel approach to use one-class support vector machines (OCSVMs) \cite{scholkopf2001estimating} to adaptively learn the normal boundary of the innovation sequence. Specifically, we train several OCSVM models using normal (i.e., non-anomalous) sensor data with different parameter values (i.e., anomaly percentages). We use the trained OCSVM models for detecting anomalies in real-time.

\subsection{One Class Support Vector Machine}

Let us define normalized innovation $\bar{\nu}$ at time $t_k$ as: 
\begin{equation}\label{eq:16}
    \bar{\nu}(t_k) = S^{-\frac{1}{2}}_k\cdot \nu_k
\end{equation}

Assume we have a training set 
$\mathcal{N}$, 
with  $l$ data points, $\bar{\nu}(t_1),...,\bar{\nu}(t_l) \in \mathcal{N}$
, sampled from a normal (i.e., non-anomalous) set. Let us define $\Phi$ as a kernel mapping function $\mathcal{N}\rightarrow \mathcal{F}$
. OCSVM solves the following quadratic program:
\begin{equation}
\label{eq:13}
\begin{aligned}
&\underset{\omega \in \mathcal{F},\boldsymbol{\xi}\in \mathbb{R}^{l},\rho\in \mathbb{R} }{\textrm{min}}\ \frac{1}{2}||\omega||^2+ \frac{1}{pl}\sum_j \xi_j - \rho \\
&\textrm{subject to}\ \ (\omega \cdot \Phi(\bar{\nu}(t_j))) \geq \rho - \xi_j,\ \xi_j \leq 0\\
\end{aligned}
\end{equation}
where $p$ is a constant parameter in the range of $(0,1)$, denoting the false positive rate of the decision boundary that classifies normal and anomalous sensor readings. Decision variables $\omega$ in model (\ref{eq:13}) define the most generalizable linear decision boundary in an infinite-dimensional space (created by the Gaussian kernel) to determine a region in the input space that encompasses at least $1-p$ percentage of data points.  Decision variables $\xi_j$ are slack variables introduced to penalize the degree of violation of the constraint $(\omega \cdot \Phi(\bar{\nu}(t_j))) \geq \rho$.

\begin{figure}[t!]
\centering
\includegraphics[width=0.48\textwidth]{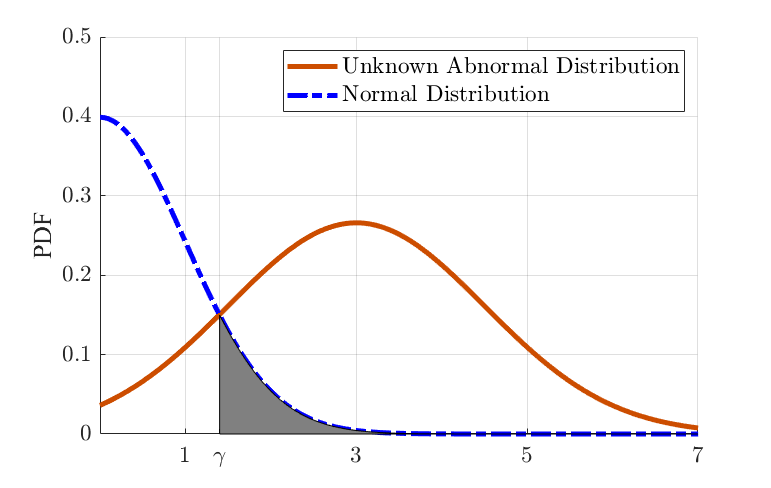}
\caption{Distribution (PDF) of normal $|\bar{\nu}(t_k)|$ and the absolute value of unknown abnormal data. The shaded area indicates false positive beyond the threshold $\gamma$.}
\label{fig:inno_abn}
\end{figure}


According to Proposition 3 in \cite{scholkopf2001estimating}, parameter $p$ provides an upper bound on the fraction of outliers in the training dataset, and asymptotically equals the fraction of outliers out-of-sample, with probability 1, under certain conditions. We train $M$ different OCSVM models, each model with a parameter $p$ selected from the set $\{p_1,p_2,...,p_M\}$ with $p_i < p_j, \forall i<j$. For a measurement sequence with dimension $m$, we compute the average of the normalized innovation sequence over a window of $N$ time intervals, up to time $t_k$, 
\begin{equation}\label{eq:17}
    \bar{\nu}(t_k)_{avr} = |\frac{1}{N}\sum_{i = k-N+1}^{k}\bar{\nu}(t_i)|_1,
\end{equation}
where $|\cdot|_1$ is the L1-norm. For small values of $\bar{\nu}(t_k)_{avr}$, i.e., when the average of normalized innovation within the current time window is small, suggesting that AEKF performs well, we choose a trained OCSVM model with large value of $p$ to ensure that we would detect even small variations and mark them as outliers. Following the same line of logic, when $\bar{\nu}(t_k)_{avr}$ is large, we choose an OCSVM model with small  $p$ to avoid unnecessary dismissal of data points where we are  not certain enough that a point is truly an outlier. 
To that end, in order to determine the parameter $p$ properly, we use the histogram of innovation values constructed from normal data to approximate the distribution of the innovation.
Without loss of generality, we assume $\nu(t_k)$ is zero mean Gaussian distributed. For the case when it is not zero mean, we first subtract the mean value, and add it back after determining the value of $p$.


The tail of the probability density function (PDF) in figure \ref{fig:inno_abn} represents drastic changes in data, and the center of the PDF represents smooth changes. The histogram of  $|\bar{\nu}|$ is an approximation of the PDF of the normal training data. As shown in figure \ref{fig:inno_abn}, parameter $\gamma$ controls the area of the shadow, and the PDF of the normal data is approximated by the histogram. We assume the absolute value of normal data as a random variable denoted as $X$ with a certain distribution and domain $D(X)$. Then, given the number of training samples $N_{train}$, the number of outliers $N_{ol}$ can be computed from 
\begin{equation}
\begin{aligned}
N_{ol} &= p\cdot N_{train}\\
&=\frac{|\{x:1-F(x) = p\}|_{mode}}{|\{x:x\in D(x)\}|_{mode}}\times N_{train} \\
&\approx\frac{|\{\bar{\nu}:|\bar{\nu}|_1 \geq \gamma\}|_{mode}}{N_{train}}\times N_{train} \\
\end{aligned}
\end{equation}
where $F(\cdot)$ is the CDF of $X$. As such, we have:
\begin{equation}
|\{\bar{\nu}:|\bar{\nu}|_1 \geq \gamma\}|_{mode} \approx p\cdot N_{train}.
\end{equation}

Finally, for $M$ OCSVM models with parameters $\{p_1,p_2,...,p_M\}$, when testing on a new data point, we have:
\begin{equation}
\begin{aligned}
\bar{\nu}(t_k)_{avr} &\in [0,\gamma_1) \Rightarrow p_1 \\
\bar{\nu}(t_k)_{avr} &\in [\gamma_1,\gamma_2) \Rightarrow p_2 \\
&...\\
\bar{\nu}(t_k)_{avr} &\in [\gamma_{M-1},\infty) \Rightarrow p_M \\
\end{aligned}
\end{equation}

In summary, assuming we have $n$ measurements, figure \ref{fig:AEKF} summaries  an implementation flowchart of the proposed algorithm, which combines AEKF and OCSVM to detect anomalies and recover the corrupt sensor readings. Specifically, at each time epoch, the following vehicle receives the measurements from both the leading vehicle and its own onboard sensors. The AEKF smooths the following vehicle's speed and location signal based on the motion model. Meanwhile the AEKF generates the innovation, which measures the discrepancy between the measurements and the prediction, and sends the innovation to the fault detector model for anomaly detection. The fault detector model consists of several OCSVM models and it can dynamically choose which one to use based on the average innovation. If there is no sensor anomaly detected, the innovation will be combined with the measurement at current time in order to generate an estimation. Otherwise, we do not trust the current sensor measurement and replace the estimation with the prediction, which will be used in the next time epoch.

\begin{figure*}[t]
\centering
\label{fig:AEKF}
\includegraphics[scale=0.53]{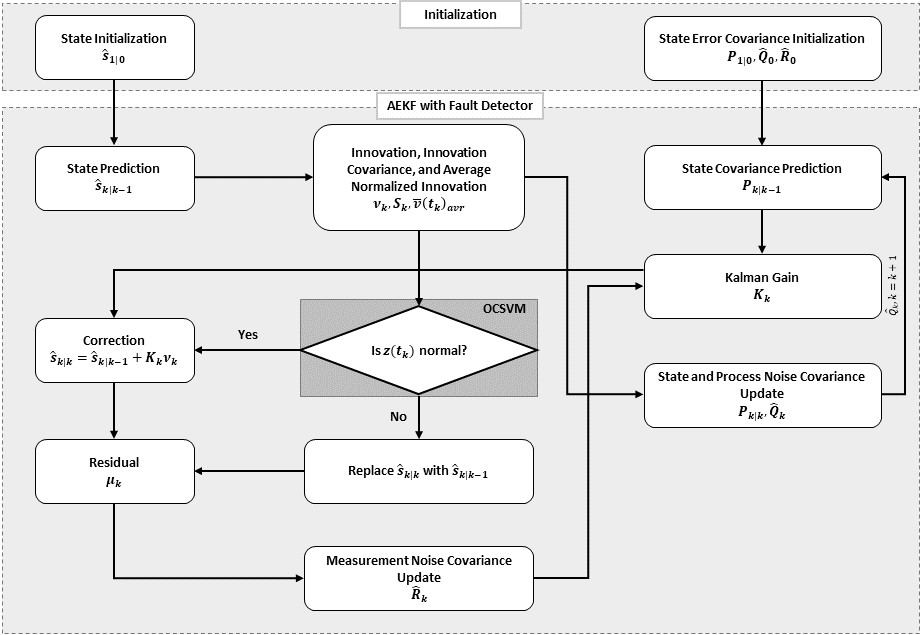}
\caption{Implementation flowchart of the proposed algorithm.}
\label{fig:AEKF}
\end{figure*}

\subsection{Anomaly Model}
The dataset for this study is generated by randomly adding anomalies to normal trajectory data, since there is no publicly available dataset on CAV trajectories that includes anomalies in sensor measurements. Specifically, we account for the four major anomaly types including short, noise, bias, and gradual drift. The detailed construction of the anomalies are as follows:
\begin{enumerate}
    \item \emph{Short:} The short anomaly type is simulated as a random Gaussian variable with mean and variance of 0 and $c_1$, respectively.
    \item \emph{Noise:} The noise anomaly type is simulated as a \textit{sequence} of i.i.d. random Gaussian variables with length of $l$, mean of 0, and variance of $c_2$.
    \item \emph{Bias:} The bias anomaly type is simulated as adding a temporarily offset to the observation. We simulate the magnitude of the anomaly as Gaussian distributed with mean of 0 and variance of $c_3$. The duration of the sequence of bias anomaly is $l$.
    \item \emph{Gradual Drift:} The gradual drift anomaly type is simulated by adding a linearly increasing/decreasing set of values to the base values of the sensors. Specifically, first we use a vector of linearly increasing values from 0 to $m$, where $m$ is a uniformly distributed random variable in range of $[0,c_4]$. The duration of the sequence is $l$. We then use a Bernoulli random variable with probability $\frac{1}{2}$ to generate one of the two outcomes of 1 and -1, by which we scale the sequence to generate increasing or decreasing drift, respectively.
\end{enumerate}
{Here} $c_i, i=\{1,2,3,4\}$ is the parameter of distribution for each type of anomaly. 

\begin{algorithm}[t]
\small
	\caption{Anomaly Generation Process}
    \begin{algorithmic}[1]
        \label{alg:1}
		\STATE $\alpha \gets$ anomaly rate; $n \gets$ number of   sensors; $\boldsymbol{\zeta} = [\zeta_1, \zeta_2,...,\zeta_n] \gets \mathcal{U}(0,1)$; $L \gets$ maximum duration of anomaly
		\FOR {\textnormal{time epoch} $t \in \mathcal{T}$}
		    \FOR {$i \in \{1,2,\dots, n\}$}
                \IF {no anomaly exists at $t$ for the $i$th sensor}
                    \IF {$\zeta_i \leq \alpha$}
                        \STATE $l \gets randi(L)$
                        \SWITCH{Anomaly type selected according to probability distribution $f_\omega$}
                            \CASE{Short}
                            \STATE Add `short' type of anomaly with parameter $c_1$
                            \ENDCASE
                            \CASE{Noise}
                            \STATE Add `noise' type of anomaly with duration $l$ with parameter $c_2$
                            \ENDCASE
                            \CASE{Bias}
                            \STATE Add `bias' type of anomaly with duration $l$ with parameter $c_3$
                            \ENDCASE
                            \CASE{Gradual Drift}
                            \STATE Add `gradual drift' type of anomaly with duration $l$ with parameter $c_4$
                            \ENDCASE
                        \ENDSWITCH
                    \ENDIF
                \ENDIF
            \ENDFOR
		\ENDFOR
	\end{algorithmic}
\end{algorithm}

We inject all four types of anomalies into each sensor reading. We assume the \textit{onset} of anomalous values in sensors occur independently. That is, we do not explicitly train OCSVM on a dataset containing interdependent sensor failures or systemic cyber attacks on vehicle sensors. However, this assumption does not preclude scenarios under which multiple sensors are under simultaneous attack.

Similar to our previous work \cite{van2019real}, we generate various datasets for our experiments with anomaly rate of $\alpha$\footnote{\href{https://github.com/next-generation-mobility-systems-lab/CAV-Sensor-Anomaly-Detection-Dataset}{https://github.com/next-generation-mobility-systems-lab/CAV-Sensor-Anomaly-Detection-Dataset}}. In addition, we simulate anomalies to start at randomly selected times, lasting for random durations (if it applies to the  anomaly type), affecting randomly selected sensors. These anomalies are then used to adjust  the corresponding sensors'  normal  readings in the original dataset, which indicate the traveling location and speed of the CAV, making them anomalous. The pseudo code describing random generation of anomalies is presented in Algorithm \ref{alg:1}. In the algorithm, $randi(L)$ denotes a discrete uniform distribution among integer numbers from 1 to $L$. Note that each anomaly type is equally likely to get selected when  an anomaly is generated.  

\section{Case Study Based On the Intelligent Driver Model}
In this section we use a well-known car-following model, namely the Intelligent Driver Model (IDM), proposed by Treiber et al. \cite{treiber2000congested}, to compare the anomaly detection performance of the traditional $\chi^2$-detector and the OCSVM model. As mentioned in \cite{treiber2014traffic}, since the IDM has no explicit reaction time and its driving behavior is given in terms of a continuously differentiable acceleration function, it describes more closely the characteristics of semi-automated driving by adaptive cruise control (ACC) than that of a human driver. However, it can easily be extended to capture the communication delay as described in the previous section. We also evaluate the impact of using the IDM motion model on anomaly detection performance. In order to evaluate system performance, we assume that the input vector containing the leading vehicle's information is not anomalous. {\color{black} When both the subject vehicle sensors and the leading vehicle's information present anomalous data, as long as the deviation of the state prediction and the measurement exceeds the threshold of the fault detector, the system still is able to detect sensor anomalies. However, when the deviation is not evident enough, e.g. under the rare circumstance when the attacker manipulates both the input vector and the measurement such that they remain consistent according to the car-following model, the fault detector will fail to detect those anomalies. However, such circumstance is not easy to occur since it requires knowledge from the part of the attacker on both the exact motion model as well as how the fault detector threshold is dynamically updated.}

Using the definition of state $s_n(t)$ and input $u_n(t)$ in the previous section, the IDM model with time delay $\tau$ can be described as the following:
\begin{equation*}
 \label{eq:21}
\begin{aligned}
\dot{x}_n(t) &= v_n(t) \\
\dot{v}_n(t) &= f_{vc}(s_n(t-\tau),u_n(t-\tau)) \\
&=a\left(1-\left(\frac{v_n(t-\tau)}{v_0}\right)^{\delta}\right.\\
&-\left.\left(\frac{s^*(v_n(t-\tau),v_n(t-\tau)-v_{n-1}(t-\tau))}{x_{n-1}(t-\tau)-x_n(t-\tau)-l_n}\right)^2\right)
\end{aligned}
\end{equation*}
with
$$
s^*(v_n,\Delta v_n) = s_0 + v_nT+\frac{v_n\Delta v_n}{2\sqrt{ab}} 
$$
where $a, b, \delta, v_0, s_0, T$ and $l_n$ are model parameters. The state vector $s_n$ and the input vector $u_n$ both have dimension of 2. For detailed information on IDM refer to \cite{treiber2000congested}. Following the typical parameter values of city traffic used in \cite{treiber2014traffic}, we set the parameter values in our study as follows: $a = 1.0, b = 1.5, \delta = 4, v_0 = 33.75, s_0 = 2, T = 1.0$, $l_n = 5$, and define the measurement function $h(\cdot)$ as:
\[h(s) = H \cdot s = 
\begin{bmatrix}
1,0\\
0,1\\
\end{bmatrix} \cdot s.\]

\begin{figure}[t!]
\centering
\includegraphics[width=0.48\textwidth]{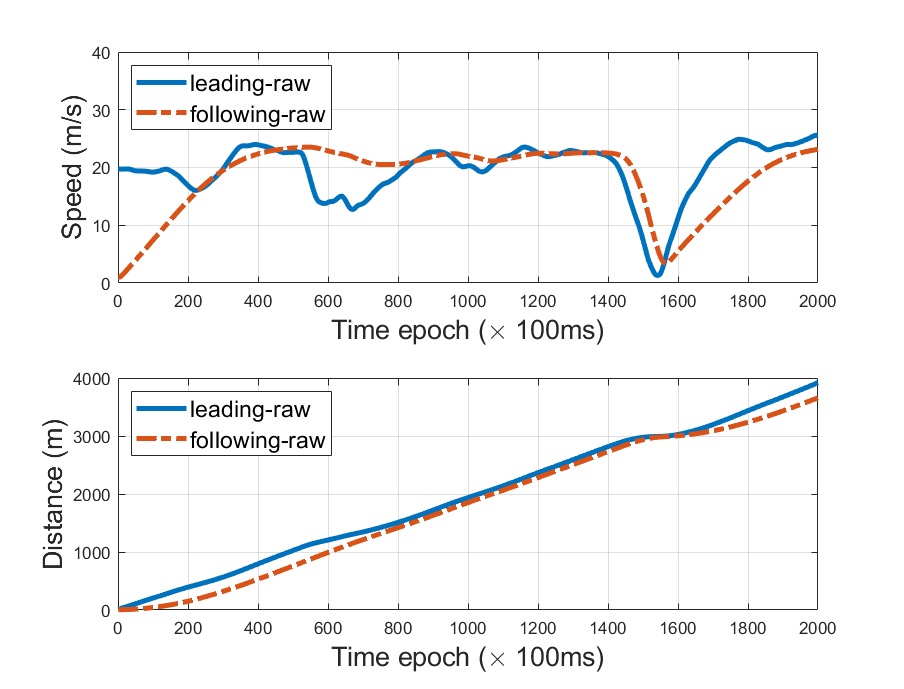}
\caption{The synthetic car-following data using SPDP data under IDM model. The first and second plot are the speed and location of two vehicles respectively.}
\label{fig:CFdata}
\end{figure}

The data for this study is obtained from the research data exchange (RDE) database constructed as part of the Safety Pilot Model Deployment (SPMD) program \cite{bezzina2014safety} funded by the US department of Transportation, and collected in Michigan. This program was conducted with the primary objective of demonstrating CAVs, with the emphasis on implementing and testing V2V and V2I communications technologies in real-world conditions. The program recorded detailed and high-frequency (10 Hz) data for more than 2,500 vehicles over a period of two years. The data features extracted from the SPMD dataset used in this study include the in-vehicle speed for one of the test vehicles with a trip length of 400 seconds (4000 samples) for training data, and 200 seconds (2000 samples) for testing data. As mentioned in section I, we assume that the vehicles are in ACC mode according to a car-following model, i.e. the IDM model. Therefore, as shown in figure \ref{fig:CFdata}, we use the extracted speed data as the leading vehicle's speed $v_{n-1}$, and generate its location $x_{n-1}$ and the following vehicle's state ($x_n$ and $v_n$) as the baseline based on the following rules:
\begin{equation}\label{eq:22}
    \begin{aligned}
        &x_{n-1}(k+1) = \ x_{n-1}(k) + v_{n-1}(k) \cdot \tau\\
        &x_n(k+1) = \ x_n(k) + v_n(k)\cdot \tau\\
        &v_n(k+1) =\\
        &\ \Delta t \cdot f\left(x_{n}(k-{\tau}),v_{n}(k-{\tau}),x_{n-1}(k-{\tau}),v_{n-1}(k-{\tau})\right) \\
        &+ v_n(k-{\tau}) + \epsilon \cdot \tau\\
    \end{aligned}
\end{equation}
where $\epsilon$ is a random term that describes the uncertainty of the following vehicle's state. In our study we generate $\epsilon$ based on a uniformly distributed random variable within the range $[-0.1, 0.1]$.
Furthermore, we add Gaussian white noise with variance 0.02 to the leading vehicle's baseline data. Since we want to test the detection performance, the noise variance should be smaller than the anomaly variance so that it would not be overpowered by the white noise. Note that adding white noise to the leading vehicle's baseline data is equivalent to using a Gaussian distributed random time delay factor of $\tilde{\tau}$ with mean $\tau$.

To demonstrate the importance of incorporating the leading vehicle's information into the following vehicle's anomaly detection procedure, we implement our framework once using the IDM car following model, and once without it, using the state-space model expressed in the following:
\begin{equation}\label{eq:23}
\begin{aligned}
    \dot{s}_n(t) &=  
    \begin{bmatrix}
        e_2^T \cdot s_n(t-\tau)\\
        0\\
    \end{bmatrix} + \theta(t)\\
    z_n(t_k) &= Hs_n(t_k)+\eta(t_k),\ k \in \{0 \cup \mathbb{Z}^+\}\\
\end{aligned}
\end{equation}
where the process noise $\theta(t)$ accounts for the introduced error.

To measure the effectiveness of our two main contributions, i.e., incorporating a car following model into the AEKF motion model and using a OCSVM fault detector, we conduct sensitivity analysis over the motion model (i.e., with and without the IDM motion model), the anomaly detection methodology (i.e., the $\chi^2$-detector and OCSVM), and the time delay (i.e., $\tau = \{0,0.5,1.5\}$). To evaluate the impact of changing models/parameters, we compute the Area Under the Curve (AUC) for each receiver operating characteristic (ROC) curve. The ROC curve is a graphical plot tool to illustrate the diagnostic ability of a binary classifier as its discrimination threshold is varied, and is created by plotting the true positive rate (sensitivity) against the false positive rate ($1-$ specificity) at various threshold settings. More specifically, we change the values of $\sigma$ in our $\chi^2$-detector, and the $\gamma$ vector for the OCSVM. Note that the vector $\gamma$ is a three-dimensional vector as we train and utilize four different OCSVM models.

The experiments are separately implemented into three scenarios, where scenario 1 contains a $\chi^2$-detector without the IDM motion model, scenario 2 contains the $\chi^2$-detector with the IDM model, and scenario 3 contains OCSVM with the IDM model. Each scenario is implemented under three experimental settings generated by varying the value of the anomaly parameter $c_i, i=\{1,2,3,4\}$. More specifically, values of 
$c_i=1$, $c_i=0.1$, and $c_i=0.05$ are used for settings 1, 2 and 3, respectively. This suggests anomalous readings become more subtle, and generally more difficult to detect, from setting 1 to setting 3. 
 Lastly, the maximum duration of anomaly, $L$, is set to 20 for each setting.
 
 Tables \ref{tab:1}-\ref{tab:3} present the AUC values of the three scenarios in our three experiment settings, with time delays of $\tau=0$, 0.5, and 1.5 seconds, respectively.
 The experiments indicate that the IDM observer-based fault detection method provides significant improvement (up to $23\%$) compared with the performance of AEKF without the IDM model, regardless of the value of time delay. Additionally, we can see that OCSVM consistently achieves a better fault detection performance than the $\chi^2$ detector.
 Results also indicate that there is a degeneracy of performance for each method as the parameter $c_i$ becomes smaller. This observation is in line with intuition, since smaller $c_i$ makes the anomaly more subtle and therefore harder to detect. Additionally, the trends of AUC values indicate that as we increase the time delay, the overall detection performance systemically deteriorates. This suggests that the time delay of the car-following model may have a negative impact on the detection performance.


\begin{table}[h]
\centering
\caption{AUC OF THREE SCENARIOS WITH $\tau = 0$ SECOND.}
\label{tab:1}
\begin{tabular}{|l|c|c|c|}
\hline
                     & $\chi^2$ without IDM &$\chi^2$ with IDM & OCSVM with IDM  \\ \hline
$c_i = 1$ & 0.9059     & 0.9723     & 0.9806     \\ \hline
$c_i = 0.1$ & 0.7764     & 0.9453     & 0.9470     \\ \hline
$c_i = 0.05$ & 0.7294  &   0.9228    &   0.9357  \\ \hline
\end{tabular}
\end{table}

\begin{table}[H]
\centering
\caption{AUC OF THREE SCENARIOS WITH $\tau = 0.5$ SECOND.}
\label{tab:2}
\begin{tabular}{|l|c|c|c|}
\hline
                     & $\chi^2$ without IDM &$\chi^2$ with IDM   & OCSVM with IDM  \\ \hline
$c_i = 1$ & 0.9024     & 0.9703     & 0.9793     \\ \hline
$c_i = 0.1$ & 0.7637     & 0.9402     & 0.9452     \\ \hline
$c_i = 0.05$ & 0.7258     &  0.9118    & 0.9260     \\ \hline
\end{tabular}
\end{table}

\begin{table}[H]
\centering
\caption{AUC OF THREE SCENARIOS WITH $\tau = 1.5$ SECOND.}
\label{tab:3}
\begin{tabular}{|l|c|c|c|}
\hline
                     & $\chi^2$ without IDM &$\chi^2$ with IDM   & OCSVM with IDM  \\ \hline
$c_i = 1$ &   0.8939   &   0.9701   &   0.9782   \\ \hline
$c_i = 0.1$ &   0.7681   &   0.9201   &   0.9294   \\ \hline
$c_i = 0.05$ &   0.7208   &   0.8875   &   0.8940   \\ \hline
\end{tabular}
\end{table}


\section{Conclusion}
\noindent This paper proposes an anomaly detection method to protect CAVs against anomalous sensor readings and/or malicious cyber attacks. We use an adaptive extended Kalman filter, informed by not only the vehicle's onboard sensors but also the leading vehicle's trajectory, in order to detect anomalous information. The well-known IDM car following model is used to incorporate the leading vehicle's information into AEKF. Lastly, to improve the anomaly detection performance, and given the fact that using AEKF the innovation is not normally-distributed, we replace the traditionally used $\chi^2$-detector with an OCSVM model. We quantify the effect of these contributions in isolation, as well as in a combined model, by conducting experiments under three scenarios: ($i$) $\chi^2$-detector without the IDM model, ($ii$) $\chi^2$-detector with the IDM model, and ($iii$), OCSVM with the IDM model. Results show that the AEKF enhanced with OCSVM and the IDM model outperforms the traditional $\chi^2$-detector-based anomaly detection used in conjunction with AEKF. Furthermore, our results indicate that a model-based anomaly detection method that can incorporate the status of the lead vehicle can further improve the detection performance. More specifically, by utilizing the leading vehicle's information to inform the AEKF and using OCSVM for anomaly detection, the proposed method can not only effectively filter out the sensor noise in CAVs, but also detect the anomalous sensor values in real-time with better performance than that without utilizing leading vehicle's information. This high performance is showcased by high AUC values in our experiments. Moreover, we study the general relationship between the delay in receiving information and the performance of anomaly detection. We show that as the time delay of signal transmission (i.e., communication channel delay or sensor delay) becomes larger, the overall detection performance deteriorates. 

The current study can be improved/expanded in multiple ways. First, the following vehicle's state and anomalous sensor values used in section IV are simulated, due to the paucity of ACC datasets with anomalies for CAVs. There exist multiple car-following datasets, but most of them were collected from human drivers. Although our study mainly focuses on the detection performance of our proposed method, it may be beneficial to directly collect ACC data from CAVs and calibrate the car-following model based on a real dataset, since the potential discrepancy between the car following model and the true traveling behaviour of the vehicle may introduce new challenges. Second, in section IV, we assume the input vector containing leading vehicle's information is not anomalous. However, our proposed method can still detect anomalies without such an assumption, i.e., as long as the discrepancy between input vector and measurement are large enough. Note that we also assume that anomalous input vector are caused either by sensor failures or false injection attacks that can be described by four types of anomaly. Third, in this study for each vehicle we only utilize a single leading vehicle's information, whereas a connected vehicle can benefit from information shared by any number of connected vehicles within its communication range as well as the infrastructure. In our future work, we plan to study the impact of incorporating multiple sources of information (e.g., multiple vehicles or RSUs) on the overall anomaly detection performance. Furthermore, we plan to expand our work to identify the source of anomaly after detection.
\bibliographystyle{./IEEEtran}
\bibliography{./IEEEabrv,Refs}

\end{document}